\documentclass[12pt]{iopart}

\usepackage{graphicx}
\usepackage{amsfonts}

\begin{document}

\title{Promoting cooperation by reputation-driven group formation}

\author{Han-Xin Yang$^{1}$ and Zhen Wang$^{2,3}$}
\address{$^{1}$Department of Physics, Fuzhou University, Fuzhou
350116, China}
\address{$^{2}$Qingdao University, Qingdao, Shandong, 266071, China\\$^{3}$Interdisciplinary Graduate School of
Engineering Sciences, Kyushu University, Kasuga-koen, Kasuga-shi,
Fukuoka 816-8580, Japan}

\begin{abstract}
In previous studies of spatial public goods game, each player is
able to establish a group. However, in real life, some players
cannot successfully organize groups for various reasons. In this
paper, we propose a mechanism of reputation-driven group formation,
in which groups can only be organized by players whose reputation
reaches or exceeds a threshold. We define a player's reputation as
the frequency of cooperation in the last $T$ time steps. We find
that the highest cooperation level can be obtained when groups are
only established by pure cooperators who always cooperate in the
last $T$ time steps. Effects of the memory length $T$ on cooperation
are also studied.
\end{abstract}


 \maketitle
\tableofcontents
\section{Introduction} \label{sec:intro}

Cooperation is widely existent in human society and animal
world~\cite{1}. Understanding the emergence of cooperation among
selfish individuals remains to be an interesting problem. So far,
evolutionary game theory has provided a powerful mathematical
framework to address this problem~\cite{2}. Researchers have
proposed various game models, among which the public goods game
(PGG) has been a prevailing paradigm~\cite{pgg}.

Due to the rapid development of complex networks~\cite{complex}, the
PGG and other evolutionary game models have been extensively studied
in various kinds of structured
populations~\cite{spatial1,spatial2,spatial3}, including regular
lattices~\cite{lattice1,lattice2,lattice3,lattice4,lattice5}, random
graphs~\cite{r1}, scale-free
networks~\cite{scale1,scale2,scale3,scale4} and dynamic
networks~\cite{dynamics1,dynamics2,dynamics3,dynamics4,dynamics5,dynamics6}.
Network structure helps cooperators survive through the formation of
clusters. Within clusters, cooperators can assist each other and
benefits of mutual cooperation outweigh losses against defector.

Apart from network reciprocity, a number of mechanisms have been
discovered that facilitate cooperation. Szab\'{o} and Hauert have
studied the voluntary participation and found that the presence of
loners leads to a cyclic dominance of the strategies~\cite{e1}.
Szolnoki and Szab\'{o} have found that inhomogeneous activity can
promote cooperation~\cite{e2}. Perc and Szolnoki have shown that
social diversity is an efficient promoter of cooperation~\cite{e3}.
Szolnoki and Perc have considered that the collective benefits of
group membership can only be harvested if the fraction of
cooperators within the group, i.e., their criticalmass, exceeds a
threshold value~\cite{e4}. Perc and Wang have found that
heterogeneous aspirations promote cooperation~\cite{e5}. Xia $et.$
$al.$ have revealed the dynamic instability of cooperation due to
diverse activity patterns~\cite{e6}. Szolnoki and Perc have found
that conformity enhances network reciprocity in evolutionary social
dilemmas~\cite{e7}. Chen and Szolnoki have found that individual
wealth-based selection supports cooperation in spatial public goods
games~\cite{e8}.

In the spatial PGG, each group is composed of a focal player and all
its nearest neighbors. Thus, for a given network, the number of
different PGG groups is equal to the network size. In the previous
studies, each PGG group is assumed to be existent all the time and
each player is always involved in several independent groups which
are determined by the interaction graph. However, in real life, not
all groups can be successfully organized and players are reluctant
to participate in some groups for various reasons. Very recently,
Szolnoki and Chen proposed a model where only those players whose
previous payoff exceeds a threshold level can establish a PGG
group~\cite{follow}. They demonstrated that a carefully chosen
threshold to establish PGG group could efficiently improve the
cooperation level.

Motivated by the pioneering work of Szolnoki and Chen, in this paper
we propose a reputation-driven group formation mechanism where PGG
groups are organized by players whose reputation reaches or exceeds
a threshold. A player's reputation is defined as the frequency of
cooperation in the past few time steps. We find that cooperation can
be greatly promoted when PGG groups are established by
high-reputation players.

\section{Model} \label{sec:methods}

Our model is described as follows.

Players are located on a $L\times L$ square lattice with periodic
boundary conditions. Each PGG group is composed of a focal player
and its four neighbors. Thus the size of each PGG group is five. A
player $i$ may participate in five different PGG groups orginized by
$i$ and its four neighbors respectively.

At each time step, every cooperator contributes a unit cost to each
involved PGG group. Defectors invest nothing. The total cost of a
group is multiplied by a factor, and is then redistributed uniformly
to all the five players in this group. We denote $i$'s strategy at
time $t$ as $s_{i}(t)=1$ for cooperation and $s_{i}(t)=0$ for
defection. At time $t$, the payoff that player $i$ gains from the
group organized by player $j$ is
\begin{equation}
\Pi_{i}^{j}(t)=-s_{i}(t)+\frac{r}{5}\sum_{x=0}^{4}
s_{x}(t),\label{2}
\end{equation}
where $x=0$ stands for player $j$, $x>0$ represent the neighbors of
$j$ and $r$ is the multiplication factor. The total payoff of the
player $i$ at $t$ is calculated by
\begin{equation}
P_{i}(t)=\sum_{j\epsilon\Omega_{i}} \Delta_{j}(t)
\Pi_{i}^{j}(t),\label{3}
\end{equation}
where $\Omega_{i}$ denotes the community of neighbors of $i$ and
itself, $\Delta_{j}(t) = 1$ if player $j$'s reputation $R_{j}(t)$
reaches or exceeds a threshold $H$ otherwise $\Delta_{j}(t) = 0$.
The reputation of a player $i$ at time $t$ is defined as the
frequency of cooperation in the last $T$ time steps, that is
\begin{equation}
R_{i}(t)=\frac{\sum_{m=1}^{T}s_{i}(t-m)}{T}.\label{4}
\end{equation}

We set $R_{i}(0)=1$ so that initially ($t=0$) all players can
organized PGG groups. When the reputation threshold $H=0$, our model
is reverted to the original model in which players can organized PGG
groups all the time. Initially, cooperators and defectors are
randomly distributed with the equal probability 0.5. After each time
step, all individuals synchronously update their strategies as
follows. Each individual $i$ randomly chooses a neighbor $j$ and
adopts the neighbor $j$'s strategy with the probability:

\begin{equation}\label{4}
W[s_{i}(t+1)\leftarrow s_{j}(t)]=\frac{1}{1+e^{[(P_i(t)-P_j(t)]/K}},
\end{equation}
where $K$ characterizes the noise introduced to permit irrational
choices~\cite{update0}.

\section{Main results and Analysis} \label{sec:main results}

We assume that players occupy nodes on a $100\times100$ square
lattice and the noise $K=0.5$. Players can be divided into four
types: $C_{s}$ ($D_{s}$) denotes cooperators (defectors) who
successfully organize PGG groups, and  $C_{f}$ ($D_{f}$) denotes
cooperators (defectors) who fail to organize PGG groups. The key
quantity for characterizing the cooperative behavior of the system
is the fraction of cooperators (including $C_{s}$ and $C_{f}$)
$\rho_{c}$ in the steady state. In all simulations below, $\rho_{c}$
is obtained by averaging over the last 5,000 time steps of the
entire 50,000 time steps. Each data is obtained by averaging over 50
different realizations.

\begin{figure}
\begin{center}
\includegraphics[width=150mm]{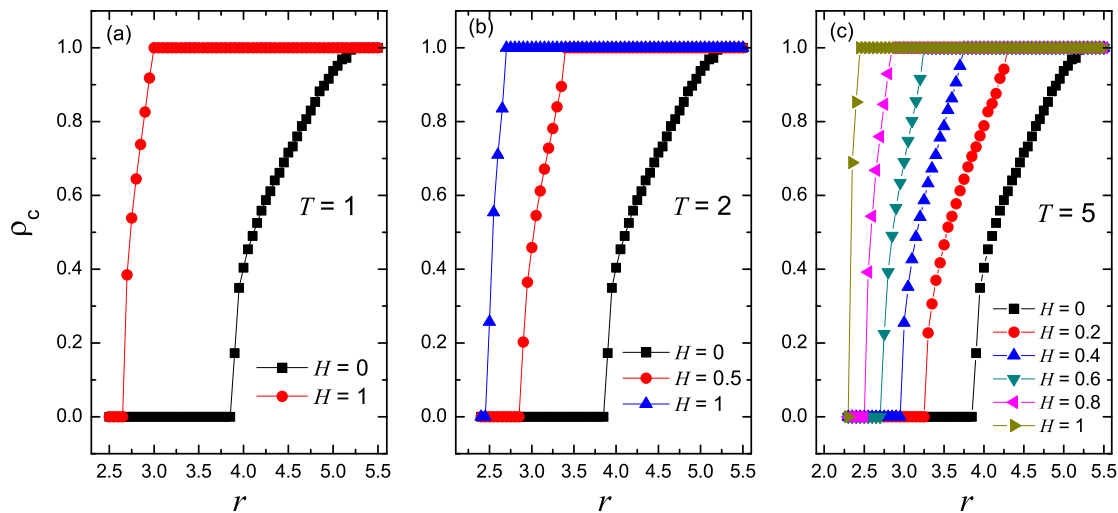}
\caption{(Color online) The fraction of cooperators $\rho_{c}$ as a
function of the multiplication factor $r$ for different values of
the reputation threshold $H$. The memory length $T$ is 2, 3, 5 for
(a), (b), (c) respectively. For each value of $T$, $\rho_{c}$
reaches the highest when $H=1$, indicating that cooperation is best
promoted when PGG groups can only be organized by pure cooperators
who always cooperate in the last $T$ time steps. }\label{fig1}
\end{center}
\end{figure}

\begin{figure}
\begin{center}
\includegraphics[width=90mm]{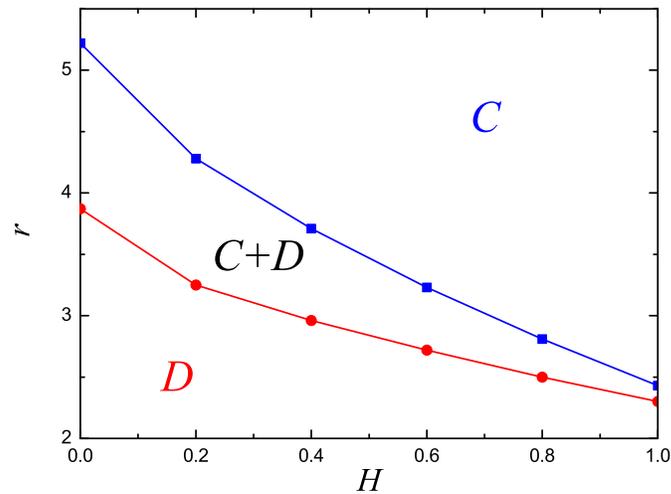}
\caption{(Color online) Full $r-H$ phase diagram for the memory
length $T=5$. There are three phases: full cooperators ($C$), full
defectors ($D$), and the coexistence of cooperators and defectors
($C+D$). As $H$ increases, the region for $C+D$ phase becomes
narrower. }\label{fig22}
\end{center}
\end{figure}

\begin{figure}
\begin{center}
\includegraphics[width=90mm]{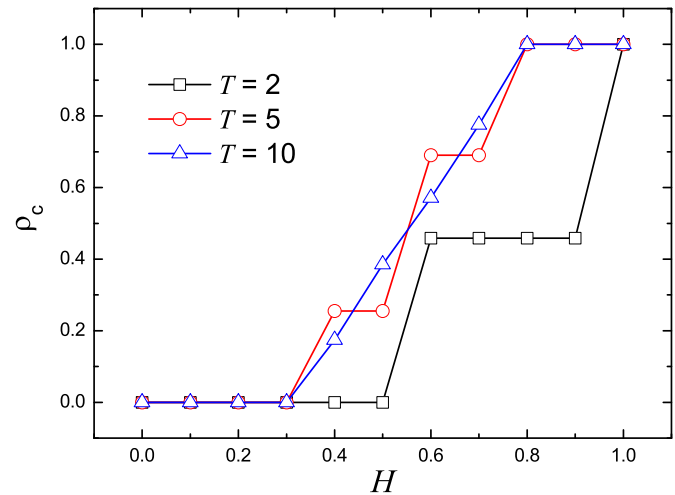}
\caption{(Color online) The fraction of cooperators $\rho_{c}$ as a
function of the reputation threshold $H$ for different values of the
memory length $T$. The multiplication factor $r=3$. For each value
of $T$, $\rho_{c}$ increases with $H$.}\label{fig33}
\end{center}
\end{figure}

Figure~\ref{fig1} shows the fraction of cooperators $\rho_{c}$ as a
function of the multiplication factor $r$ for different values of
the reputation threshold $H$. From Fig.~\ref{fig1}, we can see that
for any given value of $H$, $\rho_{c}$ increases from 0 to 1 as $r$
increases. In Fig.~\ref{fig22}, we plot the full $r-H$ phase diagram
for the memory length $T=5$. There are regions: full cooperators
($C$), full defectors ($D$), and the coexistence of cooperators and
defectors ($C+D$). One can find that the region of $C+D$ phase
becomes very narrow as the reputation threshold $H$ increases. This
phenomenon indicates that the phase transition from full $D$ to the
coexistence of $C+D$ is discontinuous when $H$ and $T$ is very
large.

\begin{figure}
\begin{center}
\includegraphics[width=90mm]{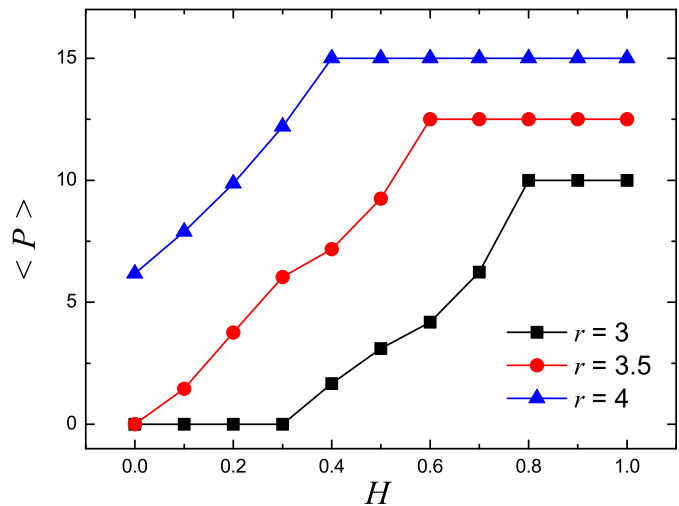}
\caption{(Color online) The average payoff of players $\langle P
\rangle$ in the steady state as a function of the reputation
threshold $H$ for different values of the multiplication factor $r$.
The memory length $T=10$. For each value of $r$, $\langle P \rangle$
increases with $H$. }\label{fig44}
\end{center}
\end{figure}

Figure~\ref{fig33} shows the fraction of cooperators $\rho_{c}$ as a
function of the reputation threshold $H$ for different values of the
memory length $T$ when the multiplication factor $r=3$. For fixed
values of the multiplication factor $r$ and the memory length $T$,
$\rho_{c}$ increases with $H$, indicating that cooperation is best
promoted when PGG groups can only be organized by pure cooperators
who always cooperate in the last $T$ time steps. Since some groups
may not be organized, it is necessary for us to investigate whether
higher cooperation level brings higher payoff. Figure~\ref{fig44}
shows the average payoff of players $\langle P \rangle$ in the
steady state as a function of the reputation threshold $H$ for
different values of the multiplication factor $r$. One can see that
for each value of $r$, $\langle P \rangle$ increases with $H$,
manifesting that the average payoff is positively related with the
cooperation level in our model.

\begin{figure}
\begin{center}
\includegraphics[width=150mm]{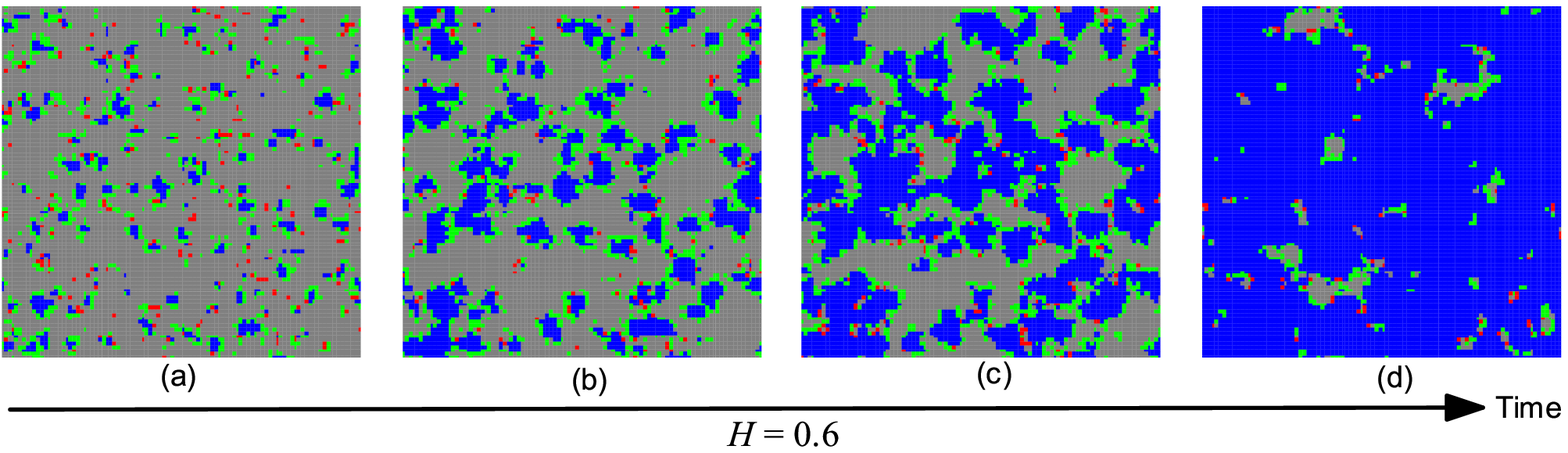}
\caption{(Color online) Snapshots of typical distributions of four
types of players at different time steps $t$ when the reputation
threshold $H=0.6$, the multiplication factor $r=3.5$ and the memory
length $T=5$. Successful cooperators ($C_{s}$) are marked by blue,
whereas failed cooperators ($C_{f}$) are denoted by green.
Successful defectors ($D_{s}$) are marked by red and failed
defectors ($D_{f}$) are denoted by gray. The time step is $t$= 5,
15, 25 and 50 for (a)-(d) respectively. In the case of high
threshold value, $C_{f}$ players form a protective layer around
$C_{s}$ clusters, which reduces payoffs of external defectors.
}\label{fig2}
\end{center}
\end{figure}

\begin{figure}
\begin{center}
\includegraphics[width=150mm]{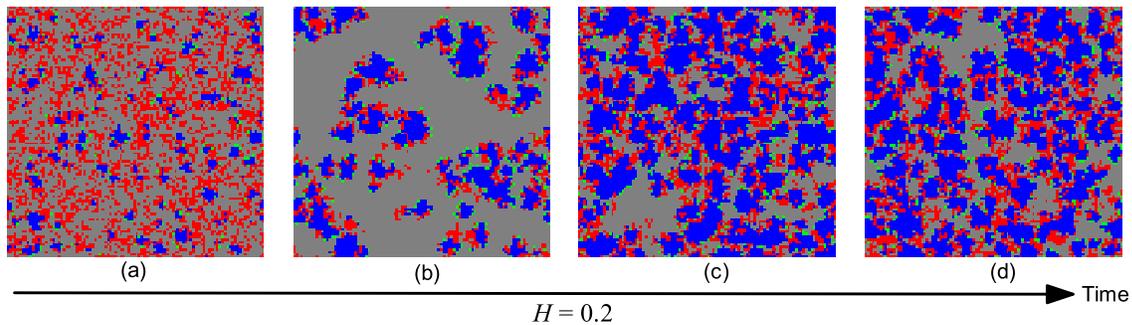}
\caption{(Color online) Snapshots of typical distributions of four
types of players at different time steps $t$ when the reputation
threshold $H=0.2$, the multiplication factor $r=3.5$ and the memory
length $T=5$. Successful cooperators ($C_{s}$) are marked by blue,
whereas failed cooperators ($C_{f}$) are denoted by green.
Successful defectors ($D_{s}$) are marked by red and failed
defectors ($D_{f}$) are denoted by gray. The time step is $t$= 5,
25, 100 and 1000 for (a)-(d) respectively. In the case of low
threshold value, $C_{s}$ clusters are surrounded by $D_{s}$ and
$D_{f}$ clusters. }\label{fig3}
\end{center}
\end{figure}

To intuitively understand the mechanism of cooperation enhancement
through reputation-driven group formation, we plot the spatial
distribution of players at different time steps for the reputation
threshold $H=0.6$ [see Fig.~\ref{fig2}] and $H=0.2$ [see
Fig.~\ref{fig3}] respectively.

For the high threshold value (e.g., $H=0.6$), most of players become
$D_{f}$ players in the early evolution [see Fig.~\ref{fig2}(a)]. As
time evolves, $C_{s}$ players form some compact clusters [see
Fig.~\ref{fig2}(b)] and $C_{s}$ clusters continually expand [see
Figs.~\ref{fig2}(c) and (d)]. Finally, $C_{s}$ players occupy the
whole system (results are not shown here). Because of the high
reputation threshold, only a very small fraction of $D_{s}$ players
can dispersedly survive during the whole evolution. It is noted that
during the evolution, $C_{s}$ clusters are surrounded by $C_{f}$
players. These $C_{f}$ players act as protective layers which can
effectively prevent the invasion of defectors. On the one hand,
defectors outside the protective layers cannot gain payoffs from
$C_{f}$ players who fail to organize PGG groups. One the other hand,
$C_{f}$ players can receive aid from $C_{s}$ players inside the
protective layers by participating in PGG groups organized by
$C_{s}$ players.

For the low reputation threshold (e.g., $H=0.2$), in the beginning
most of players become $D_{s}$ or $D_{f}$ players while $C_{s}$
players can only form tiny clusters [see Fig.~\ref{fig3}(a)]. As
time evolves, $C_{s}$ clusters expand [see Fig.~\ref{fig3}(b)].
After a long time, $C_{s}$ clusters stop expanding and are
surrounded by the three other types of players [see
Figs.~\ref{fig3}(c) and (d)]. For the low reputation threshold,
$C_{f}$ players cannot form protective layers around $C_{s}$
clusters and $C_{s}$ players are exploit by the surrounding
defectors. Thus, it becomes difficult to reach the full cooperation
in the case of low reputation threshold.

\begin{figure}
\begin{center}
\includegraphics[width=130mm]{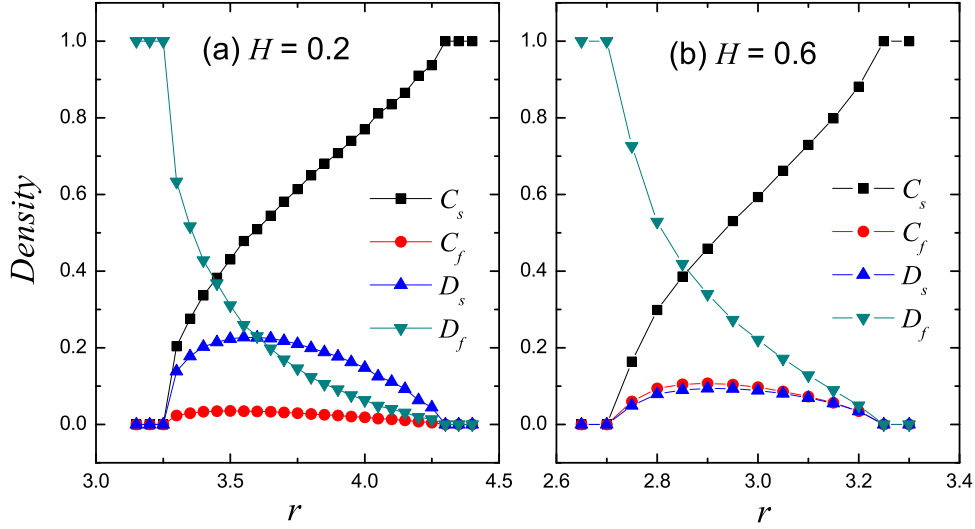}
\caption{Stationary density of the four types of players as a
function of the multiplication factor $r$ for (a) $H=0.2$ and (b)
$H=0.6$. The memory length $T=5$. $C_{s}$ ($D_{s}$) denotes
cooperators (defectors) who successfully organize PGG groups, and
$C_{f}$ ($D_{f}$) denotes cooperators (defectors) who fail to
organize PGG groups. The stationary density of $C_{s}$ ($D_{f}$)
players increases (decreases) as $r$ increases. The stationary
density of $C_{f}$ and $D_{s}$ players are maximized at moderate
values of $r$. }\label{fig4}
\end{center}
\end{figure}

\begin{figure}
\begin{center}
\includegraphics[width=150mm]{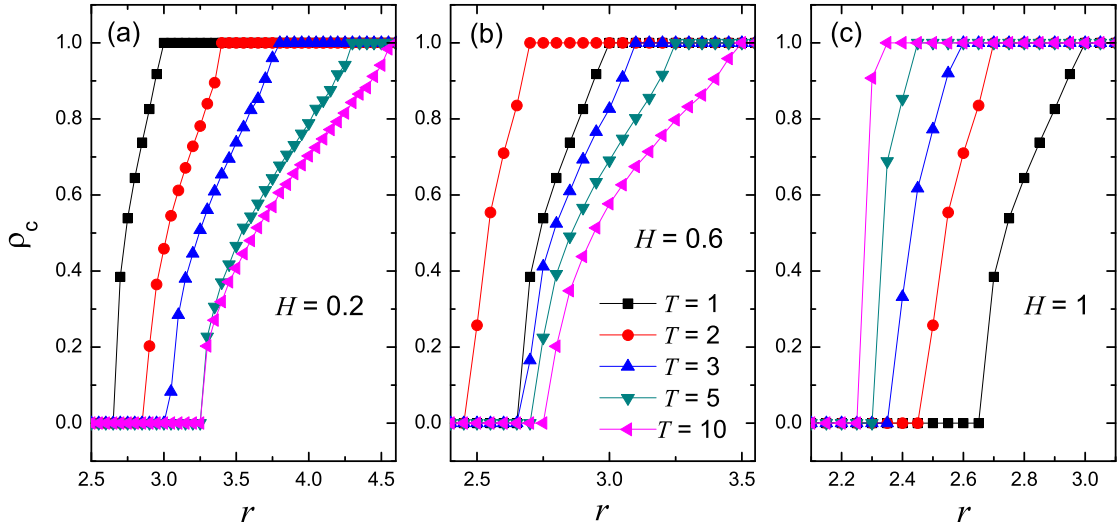}
\caption{(Color online) The fraction of cooperators $\rho_{c}$ as a
function of the multiplication factor $r$ for different values of
the memory length $T$. The reputation threshold $H$ is 0.2, 0.6, 1
for (a), (b), (c) respectively. For fixed values of $H$ and $r$, the
highest cooperation level can be reached at an optimal value of $T$
(hereafter denoted by $T_{opt}$). For $H=0.2$, $T_{opt}=1$ [see
Fig.~\ref{fig5}(a)]. For $H=0.6$, $T_{opt}=2$ [see
Fig.~\ref{fig5}(b)]. For $H=1$, the cooperation level increases as
$T$ increases, indicating that $T_{opt}=\infty$ [see
Fig.~\ref{fig5}(c)]. The value of $T_{opt}$ can be determined by the
following rule. For $i/(i+1)<H\leq (i+1)/(i+2)$, $T_{opt}=i+1$
($i$=0,1,2...). Note that for a given value of $H$ and the
corresponding $T_{opt}$, only pure cooperators who always cooperate
in the last $T_{opt}$ time steps can organized PGG
groups.}\label{fig5}
\end{center}
\end{figure}

Next, we study the stationary density of the four types of players
as a function of the multiplication factor $r$ for $H=0.2$ and
$H=0.6$ respectively. From Fig.~\ref{fig4}, one can see that the
stationary density of $D_{f}$ players decreases as $r$ increases. On
the contrary, the stationary density of $C_{s}$ players increases
with $r$. Both of the stationary density of $C_{f}$ and $D_{s}$
players are not a monotonic function of $r$. In fact, the stationary
density of $C_{f}$ and $D_{s}$ players are maximized at moderate
values of $r$. For the low reputation threshold ($H=0.2$), the
number of $D_{s}$ players is much larger than that of $C_{f}$
players [see Fig.~\ref{fig4}(a)]. For the high reputation threshold
($H=0.6$), the number of $D_{s}$ players is almost the same as that
of $C_{f}$ players [see Fig.~\ref{fig4}(b)].

Finally, we study the effects of the memory length $T$ on
cooperation. Figure~\ref{fig5} shows the fraction of cooperators
$\rho_{c}$ as a function of the multiplication factor $r$ for
different values of the memory length $T$. From Fig.~\ref{fig5}(a),
we can see that for a given value of $r$, $\rho_{c}$ decreases as
$T$ increases when the reputation threshold $H=0.2$. However, for
$H=1$, the cooperation level increases with $T$ [see
Fig.~\ref{fig5}(c)]. For $H=0.6$, the highest cooperation level is
obtained when $T=2$ [see Fig.~\ref{fig5}(b)]. In fact, we can
determine the optimal value of $T$ which leads to the highest
cooperation level as follows. For $i/(i+1)<H\leq (i+1)/(i+2)$
($i$=0,1,2...), the optimal value of $T$ (hereafter denoted by
$T_{opt}$) is $T_{opt}=i+1$. For examples, $T_{opt}=2$ when
$1/2<H\leq 2/3$ and $T_{opt}=3$ when $2/3<H\leq 3/4$ (the results
for $T_{opt}=3$ are not shown here). Note that for a given value of
$H$ and the corresponding $T_{opt}$, only pure cooperators who
always cooperate in the last $T_{opt}$ time steps can organized PGG
groups. Taking $H=0.6$ as an example, defectors with reputation
$R=2/3$ still can organize groups if $T=3$. However, for $T=1$ or
$T=2$, only pure cooperators who always cooperate in the last $T$
time steps can organized groups. According to the results in
Fig.~\ref{fig1}, cooperation can be best enhanced if PGG groups can
only be organized by pure cooperators. The criteria of pure
cooperators becomes stricter as the memory length $T$ increases.
Thus, the highest cooperation level is obtained when $T=2$.

\section{Conclusions} \label{sec:discussion}

In conclusion, we have studied the impact of reputation-driven group
formation on the evolution of cooperation. We define the reputation
of a player as the frequency of cooperation in the last $T$ time
steps. Here $T$ represents a memory length. A player can organize a
group only when his/her reputation reaches or exceeds a threshold
$H$. We find that both of the cooperation level and the average
payoff increase with the threshold $H$, manifesting that cooperation
can be best promoted when groups are only organized by pure
cooperators who never change strategy during the memory length. For
the high threshold $H$, failed cooperators who cannot successfully
organize groups form a protective layer around those successful
cooperators. The dependence of the memory length on cooperation is
found to be non-monotonic.

Our results are useful for understanding the role of reputation in
modern society. Leaders who want to organize a group should be of
high reputation. Low reputation will destroy the stability of a
group, leading to a lower cooperation level. Since individuals with
high reputation usually gain high payoff, our results are consistent
with that in Ref.~\cite{follow}. The full-cooperator state
disappears if the threshold level of payoff is too high in
Ref.~\cite{follow}. While in our model, the full-cooperator phase
still exists when the reputation threshold is very high. Together
Ref.~\cite{follow} and our work provide a deeper understanding of
the impact of group formation on the evolution of cooperation.

\section*{Acknowledge}
This work was supported by the National Natural Science Foundation
of China under Grant No. 6140308 and the training plan for
Distinguished Young Scholars of Fujian Province, China.

\end{document}